\documentclass[fleqn,usenatbib]{mnras}

\usepackage{newtxtext,newtxmath}
\usepackage[T1]{fontenc}

\DeclareRobustCommand{\VAN}[3]{#2}
\let\VANthebibliography\thebibliography
\def\thebibliography{\DeclareRobustCommand{\VAN}[3]{##3}\VANthebibliography}

\usepackage{graphicx}	% Including figure files
\usepackage{amsmath}	% Advanced maths commands
\usepackage{newtxtext,newtxmath}
\newcommand{\cqg}{Class. Quantum Grav.}

\title{Ancestral Spin Information in Gravitational Waves from Black Hole Mergers}

\author[O. Barrera and I. Bartos]{O. Barrera, I. Bartos
\\
Department of Physics, University of Florida, Gainesville, FL 32611-8440, USA}

\pubyear{2023}

\begin{document}
\label{firstpage}
\pagerange{\pageref{firstpage}--\pageref{lastpage}}
\maketitle

\begin{abstract}
The heaviest black holes discovered through gravitational waves have masses that are difficult to explain with current standard stellar models. This discrepancy may be due to a series of hierarchical mergers, where the observed black holes are themselves the products of previous mergers. Here we present a method to estimate the masses and spins of previous generations of black holes based on the masses and spins of black holes in a binary. Examining the merger GW190521, we find that assuming black hole spins that are consistent with those of merger remnants will alter the reconstructed ancestral spins when compared to results with uninformed priors. At the same time, the inclusion of black hole spins does not significantly affect the mass distributions of the ancestral black holes.
\end{abstract}

\begin{keywords}
black holes -- gravitational waves
\end{keywords}

\section{Introduction} \label{sec:intro}

When two black holes merge, they release an immense amount of energy in gravitational waves, which can be detected on Earth using gravitational wave observatories. The observation of gravitational waves by the LIGO \citep{2015CQGra..32g4001L} and Virgo detectors \citep{2015CQGra..32b4001A} revolutionized our understanding of compact objects throughout the universe as gravitational waves carry otherwise inaccessible information about black holes and their environment.

One of the most intriguing black hole mergers has been GW190521 \citep{2020PhRvL.125j1102A}. It had an unusually high reconstructed primary black hole mass of $85^{+21}_{-14}$\,M$_{\odot}$, along with high and probably misaligned spin \citep{gw190521_implications}. There are also indications that GW190521 is a highly eccentric merger \citep{GayathrieBBH,2020ApJ...903L...5R}, possibly pointing at a dynamical \citep{2021arXiv210605575G}, or AGN \citep{2022Natur.603..237S,2021ApJ...920L..42G}, origin. The high primary mass of GW190521 is difficult to explain with standard stellar evolution as pair instability in the most massive stars can rapidly disrupt the star, leaving no compact remnant behind \citep{2017ApJ...836..244W}.  

Current stellar evolution models typically predict that black holes cannot form between 50 M$_\odot$ – 135 M$_\odot$ due to either mass loss prior to core collapse, or a supernova explosion without leaving a compact remnant \cite{2017ApJ...836..244W}. Nonetheless, the boundary of this so-called mass gap is uncertain. Several studies find that stars might produce black holes up to $\sim 90$\,M$_\odot$, or even the possibility that there is no mass gap at all \citep{2018ApJS..237...13L,2020ApJ...902L..36F,2020ApJ...905L..15B,2021MNRAS.501.4514C}. The masses of GW190521 are consistent with these stellar evolutionary studies. 

If stars indeed cannot produce black holes in the mass gap, the heavy mass of GW190521 can be explained through a series of hierarchical mergers, where the observed masses are themselves the products of previous mergers (e.g., \citealt{2017PhRvD..95l4046G,2017ApJ...840L..24F,2020ApJ...900..177K,2021ApJ...915L..35K,2022ApJ...935L..26F,2022arXiv220905766M}). This dynamical creation can occur in environments with high black hole number densities such as dense stellar clusters (e.g., \citealt{2016ApJ...824L..12O,2018PhRvL.120o1101R,2019PhRvD.100d1301G,2021MNRAS.505..339M,2019MNRAS.486.5008A,2021ApJ...918L..31M}) or active galactic nuclei (e.g., \citealt{2012MNRAS.425..460M,2017ApJ...835..165B,2019PhRvL.123r1101Y,2021ApJ...908..194T,2021MNRAS.507.3362T}) (for a recent review see \citealt{2021NatAs...5..749G}).

Previously, \cite{2022ApJ...929L...1B} successfully computed the expected mass probability densities of the black holes which may have merged to form GW190521, assuming a hierarchical formation. However, the properties of ancestral black holes were investigated assuming zero spin, which may bias the obtained mass probability densities, and does not estimate the black hole spins of previous generations. 

In this paper, we expand the analysis of \cite{2022ApJ...929L...1B} to account for spins and derive ancestral mass and spin distributions. In Section \ref{sec:method} we present the method of computing ancestral properties. We discuss the application of the method to GW190521 in Section \ref{sec:results}, and conclude in Section \ref{sec:conclusion}.

%%%%%%%%%%%%%%%%%%%%%%%%%%%%%%%%%%%%%%%%%%%%%%%%%%%%%
\section{Method} \label{sec:method}

Let a remnant black hole's mass be $M$, and let $S_{\rm z}$ be its spin component that is aligned with the binary orbit, taken to be along the z axis. We constrain the discussion to aligned spins as they have substantially larger effect on the resulting distributions than spin components in the orbital plane. Let the masses and aligned spins of the parental black holes of the remnant be $m_1$ and $m_2\leq m_1$, and $s_{\rm z1}, s_{\rm z2}$, respectively. We aim to determine the probability density
\begin{equation}
p(m_1,m_2, s_{z1}, s_{z2}|p(M, S_{\rm z})).
\end{equation}
Here, $p(M, S_{\rm z})$ is the probability density of the remnant black hole's mass and spin. In the following we will use $\theta\equiv\{m_1,m_2, s_{z1}, s_{z2}\}$ to simplify notation.

%In the derivation we focus on aligned spins instead of the full spin as we found that the misaligned spin components $\{S_{\rm x},S_{\rm y}\}$ negligibly affect the ancestral parameters, and they cannot be reconstructed with meaningful accuracy. 

The probability density $p(M, S_{\rm z})$ depends on the reconstructed likelihood distribution $\mathcal{L}(M, S_{\rm z})$, which is determined using observational data, and an astrophysical prior distribution $\pi(M, S_{\rm z})$, such that 
\begin{equation}
p(M, S_{\rm z})=\mathcal{L}(M, S_{\rm z})\pi(M, S_{\rm z}).
\end{equation} 
%\cite{2022ApJ...929L...1B} previously only considered the black hole masses in determining ancestral mass distribution. Here, we further account for $S_{\rm z}$.
We compute the ancestral mass and spin distributions numerically, using the four-dimensional binned distribution $\theta_{qrtu}=\{m_{1q},m_{2r}, s_{{\rm z}1t}, s_{{\rm z}2u}\}$. The probability that the parental black holes had masses and spins $m_{1q},m_{2r}, s_{{\rm z}1t}, s_{{\rm z}2u}$, i.e. that these values fall into the $qrtu$ bin of the 4D parameter space, can be computed by marginalizing over the remnant mass and spin probability density
\begin{equation}
  p(\theta_{qrtu} \mid p(M_k, S_{\rm z})) = \sum_k\sum_l p(\theta_{qrtu} | M_k, S_{{\rm z}l}) p(M_k, S_{{\rm z}l}),
\end{equation}
where $p(M_k, S_{{\rm z}l})$ is the posterior probability of $M$ being in bin $k$ centered around mass $M_k$ with $S_{\rm z}$ being in bin $l$ centered around spin $S_{{\rm z}l}$. We can use Bayes' theorem to express the first term in the above sum as
\begin{equation}
p(\theta_{qrtu} \mid M_k, S_{{\rm z}l}) = p(M_k, S_{{\rm z}l}|\theta_{qrtu})\frac{\pi(\theta_{qrtu})}{\pi(M_k, S_{{\rm z}l})}
\end{equation}
where $\pi(\theta_{qrtu})$ is the prior probability density of $\{m_1,m_2, s_{{\rm z}1}, s_{{\rm z}1}\}$ and the prior $\pi(M_k, S_{zl})$ is 
\begin{multline}
\pi(M_k, S_{zl})= \\ \sum_q \sum_r \sum_t \sum_u p(M_k,S_{zl}|\theta_{qrtu})\pi(\theta_{qrtu}).
\end{multline}
We set the probability $p(M_k,S_{zl}|\theta_{qrtu})$ to 1 if the remnant of a binary with $\theta_{qrtu}$ masses and spins has remnant mass in the $k$ bin of $M$ and the $l$ bin of $S_{\rm z}$, and to 0 otherwise. 

We determined the conversion between parental and remnant properties, $p(M_k,S_{zl}|\theta_{qrtu})$, using the analytic solutions of \cite{2008PhRvD..78h1501T}.

%%%%%%%%%%%%%%%%%%%%%%%%%%%%%%%%%%%%%%%%%%%%%%%%%%%%%
\section{Results for GW190521}\label{sec:results}
%%%%%%%%%%%%%%%%%%%%%%%%%%%%%%%%%%%%%%%%%%%%%%%%%%%%%

\subsection{Effect of spin on mass distribution}

We used Eq. 2 to calculate the posterior probability distributions for the parental masses of both black holes in the binary GW190521. For the prior probability $\pi(M_k,S_{\rm z})$, we chose an uninformative, uniform prior that was also adopted by LIGO-Virgo \citep{2020arXiv201014527A}. We approximated integrals involving $M_k$ as Monte Carlo integrals.

The prior probability $\pi(\theta)$, was determined by taking the average of the posterior population distribution obtained from fitting the model to the GWTC-3 gravitational wave catalog (power law + peak model;  \citealt{2021arXiv211103606T}). This model considers the mass distributions for both black holes, including their correlation \citep{2018ApJ...856..173T}.

\begin{figure*}
\centering
 \includegraphics[angle = 0, scale = 0.55]{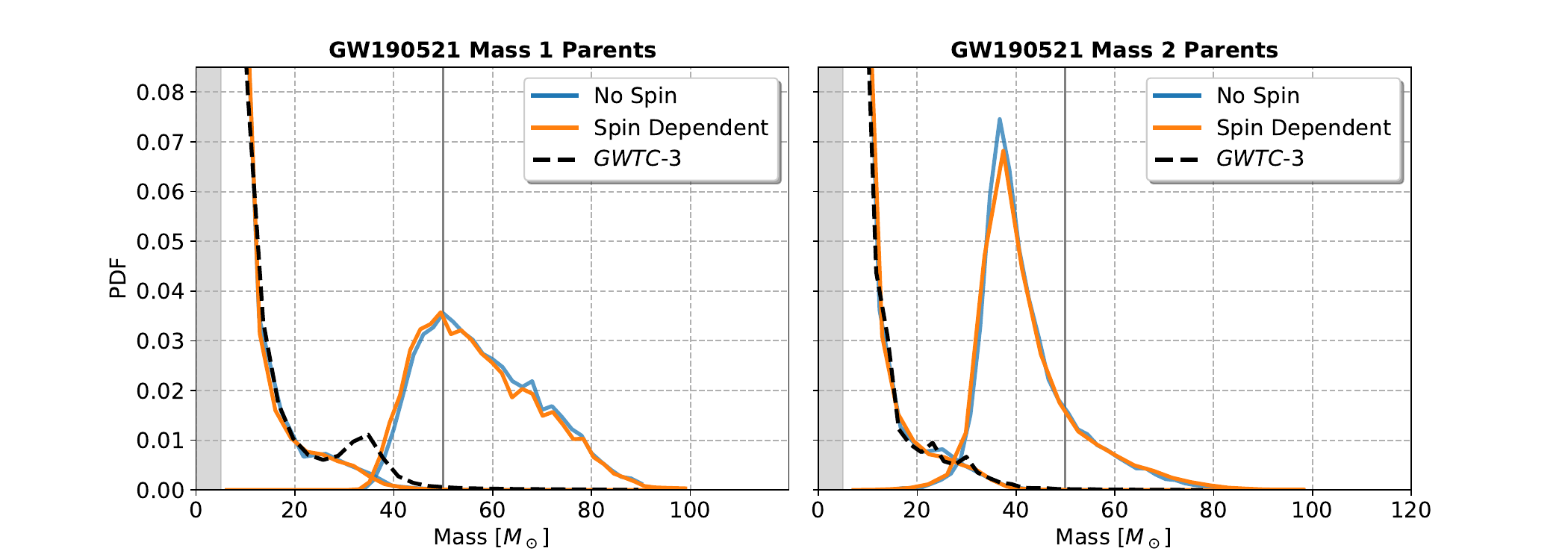}
      \caption{{\bf Probability density of ancestral black hole masses of GW190521,} separately for the more massive (left) and less massive (right) black hole in the binary. We show the corrections in probability density from a previously assumed zero spin, and for comparison the mass distribution inferred from the GWTC-3 gravitational wave catalog (see legend).
    \label{fig:massPDF}}
\end{figure*}

Results from the ancestral mass calculations are shown in Figure \ref{fig:massPDF}. To understand the role of spins, we compared the distributions found with $\{m_1,m_2, s_{{\rm z}1}, s_{{\rm z}2}\}$ against the distributions we found assuming zero spin. Our result indicates that the inclusion of spin makes no appreciable difference in the probability distribution of the ancestral masses of GW190521. 

%%%%%%%%%%%%%%%%%%%%%%%%%%%%%%%%%%%%%%%%%%%%%%%%%%%%%

\subsection{Ancestral spin}

Following the same method described above, we obtained the ancestral spin distributions of GW190521.

\begin{figure*}
\centering
 \includegraphics[angle = 0, scale = 0.55]{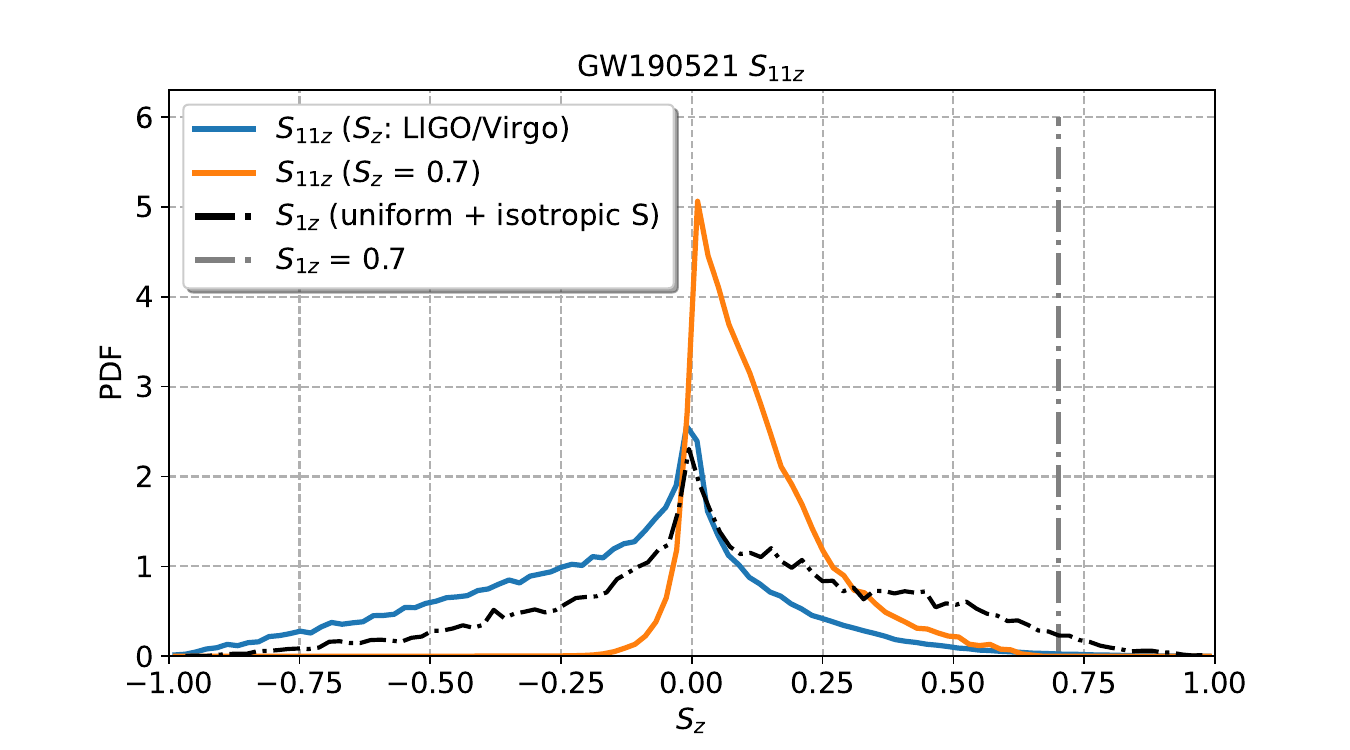}
      \caption{{\bf Probability density of ancestral black hole spins of GW190521} for the more massive black hole in the binary for different prior assumptions. We show the spin component ($S_{\rm z}$) that is aligned with the binary orbit, which can be the best constrained by observations. One prior assumption is that used by LIGO-Virgo, which considers an isotropic distribution for the orientation and a uniform distribution for the amplitude of the black hole spin $S$. Our second prior assumption is that the merger remnant has $|S|=0.7$, which is expected from the merger of two equal mass binaries. 
    \label{fig:spinPDF}}
\end{figure*}

As a first model, we adopted non-informative priors similar to those used above (see \citealt{2020arXiv201014527A}). For spin orientations, this prior is an isotropic spin distribution with uniformly distributed magnitude. The resulting ancestral spin distribution $p(s_{11z})$ for the primary parental black hole of the primary black hole in GW190521 is shown in Fig. \ref{fig:spinPDF}. We see that the resulting distribution is broad, peaking around $0$ and skewed towards negative spins. 

The above model can be improved by making use of the fact that we assume that the black hole in question, i.e. the primary of GW190521, is the remnant of a previous merger. This information will represent a different prior on the remnant black hole's spin. For a black hole merger of comparable masses, the remnant black hole's spin will be around $|S|\approx 0.7$, and will be approximately aligned with the binary orbit. To account for this we assumed $S_{1z} = 0.7$ as our prior. With this, we found a significantly different ancestral spin distribution than in the case of our uninformed spin prior. Our results are shown in Fig. \ref{fig:spinPDF}. We see that the new spin distribution $p(s_{11z})$ is now narrower than previously, and is skewed towards positive spins.
%%%%%%%%%%%%%%%%%%%%%%%%%%%%%%%%%%%%%%%%%%%%%%%%%%%%%
\section{Conclusion} \label{sec:conclusion}
%%%%%%%%%%%%%%%%%%%%%%%%%%%%%%%%%%%%%%%%%%%%%%%%%%%%%

We developed a technique to compute the probability densities of masses and spins of ancestral black holes given the probability density of the mass and spin of a remnant black hole. Our conclusions are the following:
\begin{itemize}
\item The remnant black hole's spin has negligible effect on the ancestral black hole mass probability densities.
\item Making use of the fact that merger remnants have relatively high spins aligned with the binary orbit changes the estimated spin probability density of ancestral black holes. For the case of GW190521 we found that remnant black hole mass and the assumed remnant spin carry information on the ancestral spins.
\end{itemize}
Reconstructing ancestral masses and spins will be useful for constraining the possible origin and formation site of black holes of hierarchical origin.

\section*{Acknowledgements}
We would like to thank Marek Szczepanczyk and Eric Thrane for valuable feedback. O.B. is grateful for the McNair Scholars Program, the Christopher B. Schaffer Scholarship Fund, and for the University Scholars Scholarship of the University of Florida. I.B. acknowledges the support of the Alfred P. Sloan Foundation, and NSF grant PHY-2309024. This material is based upon work supported by NSF’s LIGO Laboratory which is a major facility fully funded by the National Science Foundation. This research has made use of data obtained from the Gravitational Wave Open Science Center (https: //www.gw-openscience.org), a service of LIGO Laboratory, the LIGO Scientific Collaboration and the Virgo Collaboration. LIGO is funded by the U.S. National Science Foundation. Virgo is funded by the French Centre National de Recherche Scientifique (CNRS), the Italian Istituto Nazionale della Fisica Nucleare (INFN) and the Dutch Nikhef, with contributions by Polish and Hungarian institutes.

\section*{Data Availability}

Data generated in the course of the study is available from the corresponding author upon reasonable request.

\bibliographystyle{mnras}
\bibliography{Refs}

% Don't change these lines
\bsp	% typesetting comment
\label{lastpage}
\end{document}